\documentclass[12pt,preprint]{emulateapj}
\usepackage{amsmath}

\newcommand{\G}{\Gamma}
\newcommand{\g}{\gamma}

\slugcomment{Accepted by the Astrophysical Journal Letters}

\shorttitle{Localizing energy dissipation in blazars}
\shortauthors{Dotson, Georganopoulos, Kazanas, Pearlman}

\begin{document}

\title{A method for localizing energy dissipation in blazars using \textit{Fermi} variability}

\author{ Amanda Dotson\altaffilmark{1} \email{adot1@umbc.edu}, Markos Georganopoulos\altaffilmark{1,2}, Demosthenes Kazanas \altaffilmark{2}, Eric S. Perlman \altaffilmark{3}}

\altaffiltext{1}{Department of Physics,
University of Maryland Baltimore County, 1000 Hilltop Circle,
Baltimore, MD 21250, USA}
\altaffiltext{2}{NASA Goddard Space
Flight Center, Code 660, Greenbelt, MD 20771, USA}
\altaffiltext{3}{Department of Physics and Space Sciences,
Florida Insitute of Technology, 150 West University Boulevard,
Melbourne, FL  32901, USA}

\begin{abstract}
The distance of the \textit{Fermi}-detected blazar $\gamma$-ray emission site from the supermassive black hole is a matter of active debate.
Here we present a method for testing if the GeV emission of powerful blazars is produced within the sub-pc scale broad line region (BLR) or farther out in the  pc-scale molecular torus (MT) environment.  If  the GeV emission takes place within the BLR,  the inverse Compton (IC) scattering of the BLR ultraviolet (UV) seed photons   that produces the $\gamma$-rays takes place at the onset of the Klein-Nishina regime. This causes
 the electron cooling time to become practically energy independent and the variation of the $\gamma$-ray emission to be almost achromatic. 
 If on the other hand the $\gamma$-ray emission is produced farther out in the pc-scale MT, 
  the  IC scattering of the  infrared (IR) MT seed photons that produces the $\gamma$-rays takes place in the Thomson regime, resulting to energy-dependent electron cooling  times, manifested as faster cooling times for higher \textit{Fermi} energies.  We demonstrate these characteristics  and discuss the applicability and  limitations of our method. 
      \end{abstract}

\keywords{ galaxies: active --- quasars: general
--- radiation mechanisms: non-thermal --- gamma rays: galaxies}
\maketitle

\section{Introduction \label{section:intro}}

Blazars dominate the extragalactic $\gamma$-ray sky  \citep{abdo112fgl}, exhibiting  flares in which  the  {\sl Fermi} $\gamma$-ray flux  is seen to increase up to several times the pre-flare level \citep[e.g.][]{abdo10b}.  Because the emitting regions are unresolved, the location of the GeV emission remains unknown and this constitutes an open issue in understanding  energy dissipation in blazars.

In the leptonic model, the GeV emission of blazars is produced by IC scattering of photons off of the same relativistic electrons in the jet that give rise to synchrotron radiation \citep[for a review see][]{boettcher07}. The seed photons for IC scattering are  synchrotron photons originating within the relativistic jet (synchrotron - self Compton, SSC scattering) \citep[e.g.][]{maraschi92} or photons originating external to the jet (external Compton, EC scattering), such as  UV accretion disk photons \citep{dermer92}, UV photons originating from the BLR \citep{sikora94}, and IR photons originating from molecular torus (MT) \citep[e.g.][]{blaz00}.

Because the $\g$-ray emission of blazars can vary significantly on short times, down to the instrument limits, ranging from a few minutes (in the TeV range) \citep{aharon11} to a few hours (in the GeV range) \citep{foschini11}, light-time travel arguments constrain the size of the emitting region and it has been argued \citep[e.g.][]{tavecchio10} that this requires the emission to take place at the sub-pc scale.
Explaining  the \textit{Fermi} GeV spectral breaks seen in powerful blazars \citep{abdo10c} as a result of pair absorption on the He Lyman recombination continuum and lines, produced in the inner part of the BLR,  places the GeV emission well within the  BLR  \citep{poutanen10}.  Very short distances $ \lesssim 10^{16}$ cm), however, where accretion disk photons dominate are not favored, because of the high pair production $\gamma$-ray opacity  \citep[e.g.][]{ghisellini09}.

Gamma-ray flares have  been associated with  radio and optical flares,  placing the GeV emission outside the BLR: \cite{lahteenmaki03}, using the temporal separation of the radio and $\gamma$-ray fluxes, argued that the $\gamma$-ray emission site is at a distance of $\sim 5$ pc  from the central engine.  Connections have been also  drawn between $\gamma$-ray outbursts, optical flare polarization changes, and ejections of superluminal radio knots on the pc scale:
\cite{marscher10} associated short $\gamma$-ray variations with the passage of disturbances  from the radio core,   parsecs away from the black hole. In the case of OJ 287,  \cite{agudo11} found $\g$-ray flaring to be simultaneous with mm outbursts associated with a VLBA jet component 14 pc downstream from the radio core.

 Here we propose a method to distinguish if the GeV emission of powerful blazars takes place within the sub-pc BLR or further out within the pc-scale MT. In \S \ref{cooling} we discuss our understanding of the BLR, MT, and SSC photon fields  and where each one dominates. We show that cooling on the UV BLR photons results in achromatic cooling times in the \textit{Fermi} band, while cooling on the MT IR photons results to faster cooling times at higher $\gamma$-ray energies. In \S \ref{diagnostic} we present our method, discussing the dilution of an energy-dependent cooling time from  light travel time effects and discuss the applicability of our method. Finally, in \S \ref{conclusions}  we  present our conclusions.

\section{Cooling  of GeV emitting electrons}
 \label{cooling}

\subsection{Characterizing the BLR and the MT}
\label{sec:characterizing}

Reverberation mapping of the BLR points to a typical size of $R_{BLR} \approx 10^{17} \, L_{d,45}^{1/2}$ cm  \citep[e.g.][]{kaspi07,bentz09},
where $L_{d,45}$ is the accretion disk luminosity in units of $10^{45}$ erg s$^{-1}$.
Because of this scaling, and assuming that $L_{BLR}=\xi_{BLR} L_d$, where $\xi_{BLR}\sim 0.1$ is the BLR covering factor,  the energy density inside the BLR  ($U_{BLR} \sim 10^{-2} \, \mathrm{erg \; cm^{-3}}$)  is not expected to vary widely from object to object \citep{ghisellini09}.
As was shown by \cite{tavecchio08}, when transformed to the jet frame, the BLR  spectrum, dominated by $H$ Ly$\alpha$ photons, is broadened and boosted in luminosity due to relativistic effects and  can be well-approximated by a blackbody spectrum that peaks at $\nu_{BLR} = 1.5\, \Gamma \,\nu_{Ly\alpha}$ ($\epsilon_{BLR} \approx  3 \times 10^{-5}\, \G)$, where $\G$ is the bulk Lorentz factor and $\epsilon$ is the photon energy in mc$^2$ units.       
	
	In the MT the dust temperature ranges from 
 $T\sim 300$ K \citep{landt10} ( to $T\sim1200$ K \citep{cleary07,malmrose11}, with hotter emitting material expected to be  closer to the central engine \cite[e.g.][]{nenkova08}. 
 Because of the larger size of the MT, it has been probed by reverberation mapping only 
for Seyfert galaxies  \citep[e.g.][]{sug06} which have lower luminosities and therefore smaller MT
sizes.   These studies, along with near-IR interferometric studies
 \cite[e.g.][]{kis11}  also support an $R\propto
L_d^{1/2}$ scaling, suggesting that $U$ is  similar for sources of
different luminosities. 
 Adopting the scaling  of \cite{ghisellini09}, 
 $R_{MT}=2.5 \times 10^{18} L_{d,45}^{1/2}$ cm and assuming a covering factor $\xi_{MT}\sim 0.1-0.5$, the  MT photon energy density
 is 
$U_{MT} \sim 10^{-4}$  erg  cm$^{-3}$.
We also adopt a  temperature  $T_{MT} \approx 1000$, corresponding to a blackbody peak at $\nu_{MT} \approx 6.0 \times 10^{13}$ Hz ($\epsilon_{MT} \approx 5 \times 10^{-7}$).

\subsection{Which seed photons dominate where?}

For an isotropic photon field, the co-moving (jet frame) energy density $U'$  is  $U'\approx (4/3)\Gamma^2 U$, while for  photons entering the emitting region from behind,  $U' \approx  (3/4)\Gamma^{-2}U$ \citep{dermer94}.  
	If the emission site is located at $R \lesssim R_{BLR}$, the BLR photon field can be considered isotropic in the galaxy frame and, using the BLR energy density discussed in \S \ref{sec:characterizing}, $U'_{BLR} \sim \,1.3\times \; \Gamma_{10}^{2}  \;  \mbox{erg cm}^{-3}$. Similarly, the  MT photon field is isotropic inside the BLR and its co-moving seed photon energy density is $U'_{MT}\sim  1.3 \times 10^{-2} \; \Gamma_{10}^2$  erg cm$^{-3}$. Clearly, at  $ R_{BLR}$ scales, $U'_{BLR}$    dominates over $U'_{MT}$  by a factor of $\sim 100$. 
 If the emission site is located at $R \sim R_{MT}$,  then the BLR  photons enter the emitting region practically from behind, so that  $U'_{BLR}  \sim 7.5\times 10^{-7}  \Gamma_{10}^{-2}$ erg cm$^{-3}$.  The MT photons  retain the same co-moving energy density  as before and, in this case,  the MT  dominates the co-moving photon energy density.
  
 	These external photon field co-moving energy densities need to be compared to the synchrotron one. If $R_{blob}$ is the size of the emitting region, then the  co-moving synchrotron photon energy density is $U'_{S}  \approx { L_{s} / (4\pi c R_{blob}^2 \delta^4)}$, where $\delta$ is the usual Doppler factor and $L_s$ the synchrotron luminosity.  An upper limit to $R_{blob}$ is set by the variability timescale $t_v$: $R_{blob}\lesssim c t_v\delta/ (1+z)  $, where $z$ is the source redshift.  We then obtain a lower limit  $U'_s \approx  L_s (1+z)^2/( 4\pi c^3 t_v^2\delta^6)$ or
$U'_s  \sim 6.3 \times 10^{-2} L_{s,46} (1+z)^{2} t_{v, 1d}^{-2}   \delta_{10}^{-6} \;  \mbox{erg cm}^{-3}$, where $t_{v, 1d}$  is the  variability time in days.   
Setting $\delta=\Gamma$	the condition $U'_s<U'_{ext}$ requires
	\begin{equation}
	\label{eq:gamma}
	\Gamma \gtrsim   \left(3 L_s (1+z)^2\over16 \pi c^3 t_v^2 U'_{ext}\right)^{1\over 8}=8.6 \left({L_{s,46} (1+z)^2 \over t_{v,1d}^2 U'_{ext,-4}}\right)^{1\over 8}.
	\end{equation}  
	
This translates to  $\Gamma\gtrsim 4.8\times  L_{s,46}^{1/8}(1+z)^{1/4} t_{v,1d}^{-1/4}$	for the emission site at $R_{BLR}$ scales and  
	$\Gamma\gtrsim 8.6\times  L_{s,46}^{1/8}(1+z)^{1/4} t_{v,1d}^{-1/4}$ for the emission site at $R_{MT}$ scales.  In general, powerful quasars exhibit superluminal motions requiring $\Gamma\sim 10-40$ \citep{jorstad05, kharb10} and therefore, the dominant source of seed photons for these sources should be the BLR or the MT, provided the $\gamma$-ray emission site is at a distance not significantly larger than the size of the MT.  A recent finding by \cite{meyer12} that the ratio of $\gamma$-ray to synchrotron luminosity increases with increasing radio core, as expected for EC emission \citep{dermer95,georganopoulos01}, supports further the possibility that powerful blazars are EC emitters.

\subsection{Cooling in the BLR vs Cooling in the MT}

 {\sl The critical difference between the BLR and the MT is the energy of the seed photons:  the BLR produces UV photons while the MT produces IR photons.  This difference by a factor of $\sim 100$ in typical photon energy is critical in that it affects the energy regime in which the GeV-emitting electron IC-cooling takes place, and thus the energy dependence of the electron cooling time.}

For this consideration to be relevant, IC cooling must dominate over synchrotron cooling.
This condition is satisfied, since in powerful blazars the IC luminosity clearly dominates over the synchrotron luminosity, reaching  in high states an IC luminosity higher that the synchrotron one  by a factor of up to $\sim100$ \citep[e.g.][]{abdo10a,meyer12}.

For electrons cooling in the Thomson regime ($\gamma\epsilon_0 << 1$, where both the electron Lorentz factor $\gamma$ and the seed photon energy $\epsilon_0$ are measured in the same frame) the cooling rate $\dot{\gamma} \propto \gamma^2$.  For electrons with $\gamma \epsilon_0 >>  1$ cooling takes place in the Klein-Nishina (KN) regime with $\dot{\gamma} \propto \ln \gamma$ \citep{blum70}.  For the broad intermediate regime, $ 10^{-2}\lesssim \gamma\epsilon_0\lesssim 10^2$,  a parametric approximation has been suggested \citep{moderski05}. In what follows we calculate numerically the electron energy loss rate $\dot{\gamma}$, which for monoenergetic seed photons  of number density $n_0$ and dimensionless energy $\epsilon_0$ is 

\begin{equation}
	\label{eq:gdotint}
	\dot{\gamma} = n_0 \int_0^\infty  \frac{3 \sigma_T c}{4 \epsilon_0}f(x) (\epsilon - \epsilon_0) \mathrm{d}\epsilon,
\end{equation}
where, following  \cite{jones68},
\begin{equation}
	\label{eq:fx}
	f(x)= 2x\log x + x + 1 - 2x^2 + \frac{\left(4\epsilon_0\gamma x\right)^2}{1+4\epsilon_0 \gamma x}\left(1-x\right)
\end{equation}
with
$x= \epsilon/[4 \gamma^2 \epsilon_0 \left(1 - \epsilon/\gamma\right)]$.
	
\begin{figure}
	\epsscale{1.3}
	\plotone{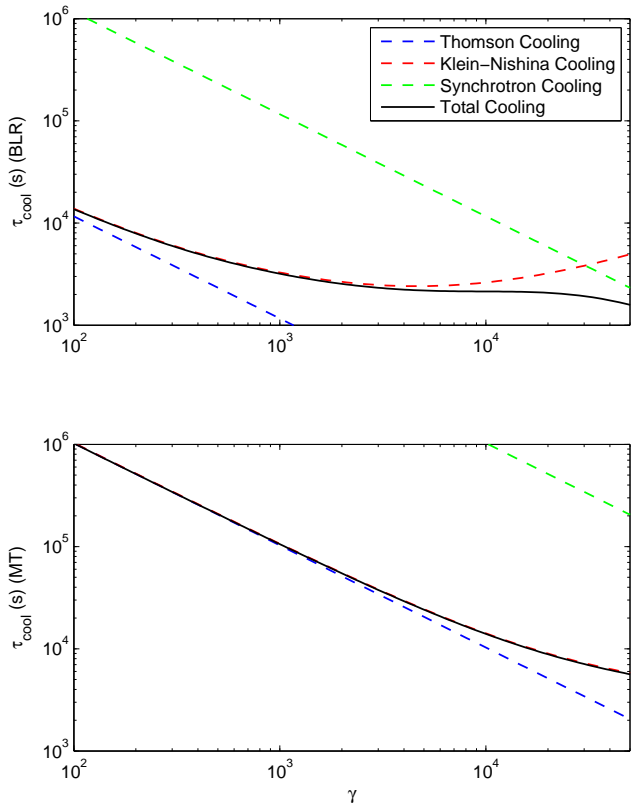}
	\caption{Electron cooling time in the galaxy frame as a function of  $\gamma$. Top panel: Blazar emission site located in the BLR. Bottom panel: Blazar located in the MT.  Broken lines represent the various cooling mechanisms and the solid black line is the total cooling time.  Plots were calculated for seed photon energies $\epsilon_{0,BLR} = 3 \times 10^{-5}$ and $\epsilon_{0,MT} = 5 \times 10^{-7}$, seed photon energy densities $U_{BLR} =  10^{-2}$ erg cm$^{-3}$, $U_{MT} =  10^{-4}$ erg cm$^{-3}$ and  magnetic field energy $U_B=100\,  U_{EC}$.}
	\label{fig:blrmt}
\end{figure}

	The effects of the transition between Thomson and KN regimes on the electron energy distribution (EED) and the resultant spectrum of the synchrotron and IC emission have been studied before \citep[e.g.][]{blum71,zdz89,dermer02,sok04,mod05,kusun05,georganopoulos06,manolakou07,sikora09}. 		

In short, the transition from Thomson to KN cooling in the EED can be seen for the steady state $n(\gamma) \propto \int Q(\gamma) d\gamma/\dot{\gamma}$: for a power law injection $Q(\gamma) \propto \gamma^{-p}$, the cooled EED hardens from $n(\gamma) \propto \gamma^{-(p+1)}$ in the Thomson regime to $n(\gamma) \propto \gamma^{-(p-1)}$ in the KN regime.  The transition takes place gradually at $\gamma \sim \epsilon_0^{-1}$.  A second transition back to $n(\gamma) \propto \gamma^{-(p+1)}$ is expected at higher  $\gamma$, as synchrotron losses (assumed to be less significant than Thomson EC losses) become larger than KN losses.  The imprint of these considerations on the observed synchrotron and IC spectra has been considered for the case of the X-ray emission of large scale jets \citep{dermer02}  and in the case of blazars \citep{kusun05,georganopoulos06,sikora09}.

		Spectral signatures, however, are not unique  and can be diluted or altered by extraneous causes (e.g. the accelerated electron distribution has intrinsic features that deviate from a power law, the synchrotron emission is contaminated by the big blue bump, the IC emission is modified by pair production absorption from synchrotron or external photons). For these reasons their use in understanding where the GeV emission takes place is problematic. \\

\section{Locating the GeV emission site}
\label{diagnostic}

		There is however another observable aspect of the Thomson - KN transition that is free of the issues faced by relying on spectral signatures:
{\sl the energy dependence of the electron cooling time can be used to evaluate the regime at which the electrons that produce the GeV radiation cool, and through this evaluate where the emission takes place.}

	In the Thomson regime, the electron cooling time $t_c=\gamma/\dot{\gamma}$ scales as $\gamma^{-1}$, while in the KN regime the electron cooling time scales as as $\gamma/\ln\gamma$.  These two regimes connect smoothly forming a wide  {\sl valley} around $\gamma \epsilon_0\sim 1$ in which the cooling time is practically energy-independent.  
	The energy of the seed photons originating in the BLR is greater than that of photons from the MT by a factor of $\sim 100$.  As a result, (see Fig. \ref{fig:blrmt}) the location of the energy-independent valley of the BLR case is manifested at electron energies lower by a factor of $\sim 100$ than that of the MT case.
   In turn, the energy dependence of $t_c$ causes a related energy dependence in the observed falling time $t_f$ of the flare. The falling time $t_f$ of \textit{Fermi} light curves can be used, therefore,  to determine whether a flare occurs within the BLR or within the MT.

\subsection{Flare decay in the BLR vs flare decay  in the MT}

	We utilize an one-zone code  to demonstrate the effect of a flare occurring within the BLR versus a flare occurring within the MT.
	The code assumes the injection of a power law electron distribution and follows its evolution taking into account the KN cross section, SSC losses,  and assuming a black body external photon field. The implicit numerical scheme is similar to that used by \cite{chiaberge99} and \cite{graff08}. After the code reaches a steady-state, an increase in the injection rate is introduced that lasts for a short time before it returns to its initial value.  
	The code was initialized to simulate the flare occurring within the BLR and the MT respective, described by the values discussed in \S \ref{cooling}.  We assumed a Compton dominance $L_{IC,max}/L_{synch,max} \approx 50$,  source size $R=3\times10^{16} \mathrm{cm}$, bulk Lorentz factor $\Gamma = 10$.  Deep in the Thomson regime (MT case)  the Compton dominance corresponds to $U_{EC}/U_{B}$, where $U_{EC}$ is the external energy density and $U_B$ is the magnetic energy density. However, as cooling transitions from the Thomson to the KN regime (BLR case), the ratio $U_{EC}/U_B$ over-estimates the observed Compton dominance by a factor of $\sim 2.$

\begin{figure}
\epsscale{1.3}
\plotone{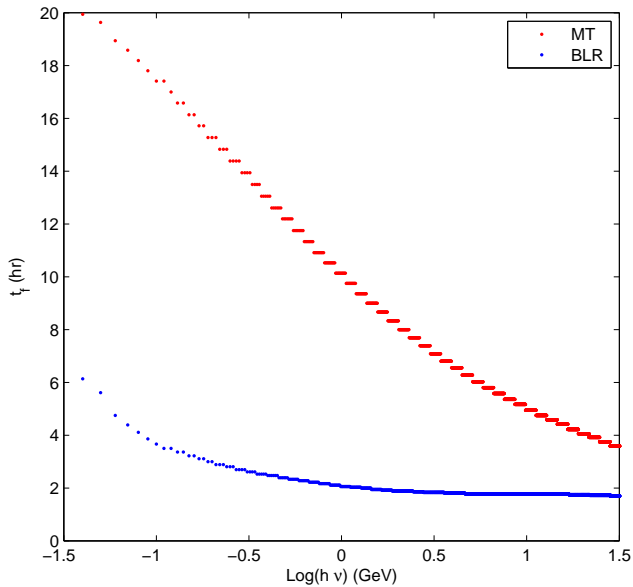}
	\caption{$1/e$ falling time in the galaxy frame versus observing energy (GeV) for MT seed photons (red), BLR seed photons (blue)}
	\label{fig:epscool}
\end{figure}

\begin{figure}
\epsscale{1.3}
\plotone{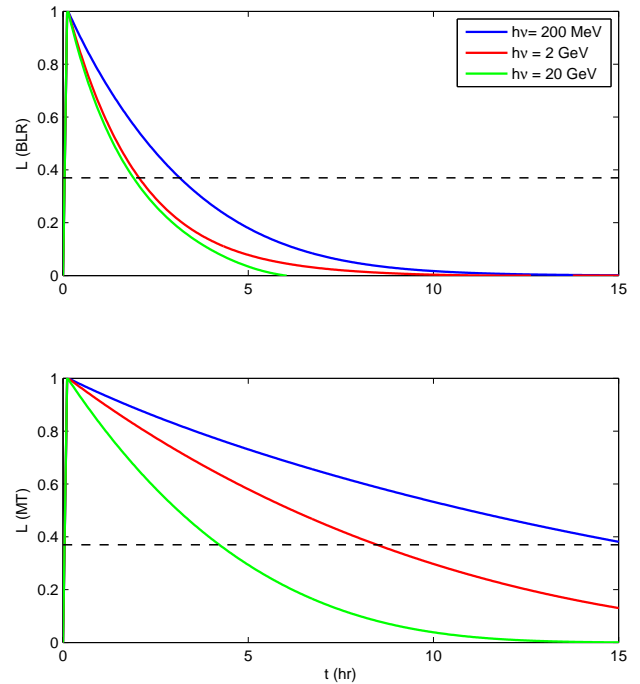}
	\caption{Light curves at $200$ MeV, $2$ GeV, and $20$ GeV for BLR seed photons (upper panel) and MT seed photons (lower panel), normalized for a steady state L=0 and maximum L=1 (in arbitrary units). Times are in the galaxy frame. The dashed black line is the $1/e$ decay time. Times are in the galaxy frame.}
	\label{fig:flarelc}
\end{figure}

In the BLR case (Fig. \ref{fig:epscool}, blue curve) IC cooling occurs at the onset on the KN regime, resulting in a near-flattening of $t_f$ within the {\sl Fermi} band, where a change in energy by a factor of 100 (from 200 MeV to 20 GeV) results in a change in cooling time of less than an hour, with all the {\sl Fermi} energies having practically indistinguishable decay times (between $\sim 2-3$ hours). 
In the MT case (Fig. \ref{fig:epscool}, red curve) cooling occurs mostly in the Thomson regime and $t_f$ is heavily energy-dependent; the same change in energy by a factor of 100 results in a change in cooling time by $\sim 10$ hours.
 This is also apparent in the light curves shown in Fig \ref{fig:flarelc}.
	A flare occurring within the BLR (Fig \ref{fig:flarelc}, top panel) has comparable values of $t_{f}$ within the {\sl Fermi} energy range.   A flare occurring within the MT exhibits distinct falling times that decrease with observing energy (\ref{fig:flarelc}, lower panel).

	{\sl For sufficiently bright flares with short decay times (comparable to the cooling times of few to several hours anticipated), {\sl Fermi} light curves can be generated at multiple energies, and the energy dependence (or lack thereof) of $t_{f}$ can be used to reveal the location of the GeV emitting site.}
	
\subsection{Light Travel Time Effects}
\begin{figure}
	\label{fig:lctime}
	\epsscale{1.3}
	\plotone{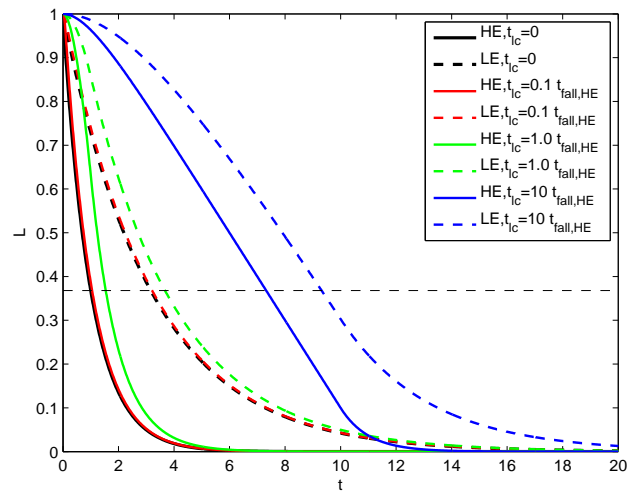}
	\caption{Exponentially decaying light curves of a flare in the MT. High and low energy luminosities  ($\epsilon_{HE}/\epsilon_{LE} = 10$) are plotted (normalized to 1) for increasing $t_{lc}$.  Solid lines represent $\epsilon_{HE}$  and dashed lines $\epsilon_{LE}$. Colors represent  $t_{lc}$ at differing factors of $t_f$, where $t_{f, HE}$ is normalized to 1. The dashed black line is the $1/e$ decay time.}
	\end{figure}

	Because the GeV emitting region is not a point source, the observed $1/e$ decay time $t_{obs}$ of the light curve is a convolution of  $t_f$  and the blob light-crossing time $t_{lc}$.  Any other variations  (e.g. a gradual decrease of the electron density) will have a similar effect. For BLR achromatic cooling, we expect the light curves to remain achromatic after considering light crossing time effects.  	
	 To see the effects of $t_{lc}$ in the case of MT cooling, where, $t_f \propto \epsilon^{-1/2}$, we consider a source with exponentially decaying emission coefficient, $j(t) = j_0 \exp^{-t/t_f}$, and calculate the expected light curves for two energies differing by 10 and for a range of $t_{lc}$. As can be seen in  Fig \ref{fig:lctime}, although $t_{obs}$ increases with increasing $t_{lc}$, the difference of $t_{obs}$  between $\epsilon_{LE}$ and $\epsilon_{HE}$ is practically preserved  for all $t_{lc}$. Moreover, because for given	 $\epsilon_{LE}$ and $\epsilon_{HE}$  the ratio of $t_f$ is also given,  observations of the $1/e$ decay times for $\epsilon_{LE}$ and $\epsilon_{HE}$  can be used to find both $t_{lc}$ and  $t_f$ for each energy.

  \subsection{The feasibility of the diagnostic}
  
  Very bright flares, such as the November 2010 flare of 3C 454.3 
\citep{abdo3c454}
have enough photon statistics to produce light curves in two
{\sl Fermi } energy bands. The practical application of our diagnostic
requires that the decay times in different energy bands have 
statistically meaningful differences if the flare occurs within the MT.
To evaluate if this is the case for a flare as bright as that of  3C
454.3, we
adopted  the maximum observed errors in the $0.1 - 1$ GeV range and $> 1$
GeV range   \citep{abdo3c454} and produced simulated data assuming
energy-dependent cooling  in the form of exponentially decaying light
curves.
We fitted  an exponential decay function to the simulated data for each
energy band, with the characteristic decay time  as a free parameter.  
The decay times we recovered were statistically distinguishable,
suggesting that our diagnostic is applicable, at least for the brightest
flares observed.

\subsection{Constraints from  upper limits of the decay time difference}

For flares from which no statistically significant difference in the decay
time can be established, constraints can still be imposed on the location
of the {\sl Fermi} detected emission by considering $\Delta t_{max}$, the
maximum  decay time difference allowed by the data, between a high energy
$\epsilon_{HE}$ and a low energy $\epsilon_{LE}$.
The requirement that $\Delta t_{max}\geq
t_f(\epsilon_{LE})-t_f(\epsilon_{HE})$ can be casted as upper limit for
the location $R$ of the {\sl Fermi} emission
\begin{equation}
R   \lesssim  \left[
{2 \sigma_T L_{MT} \Gamma^2 \Delta t_{max}
\over
3\sqrt{3} \,\pi m_e c^2 (1+z)^{1/2}\epsilon_{MT}^{1/2}  
(\epsilon_{LE}^{-1/2} - \epsilon_{HE}^{-1/2}) }
\right]^{1/2},
\end{equation}
where $L_{MT}$ is the MT luminosity.
In terms of quantities used in \S \ref{sec:characterizing} and for
$\epsilon_{LE}, \; \epsilon_{LE}$ corresponding to 100 MeV and 1 GeV,
$ R \lesssim 2.3 \times 10^{18}\; \Gamma_{10} (\Delta t_{max,h}
L_{MT,45}/(1+z)^{1/2})^{1/2} \mbox{cm}$. The Lorentz factor can be estimated
from VLBI studies \cite[e.g.][]{jorstad05} and the MT luminosity can be
taken to be a fraction $\xi\sim 0.1-0.5$ of the accretion disk luminosity
or in some cases can be directly measured \cite[e.g.][]{landt10,
malmrose11}.

	\section{Conclusions}
\label{conclusions}

	We  presented a diagnostic test for determining if the GeV emission
in powerfull blazars originates from within or outside the BLR. Our
method utilizes the fact that if electrons cool via IC scattering on BLR
photons, cooling occurs at the onset of the KN regime and the resulting
cooling times (and thus light curves) should be achromatic. If cooling occurs via IC
scattering of less energetic MT photons, the electron cooling times, and
thus light-curves, should exhibit significant energy-dependence and
decrease as photon energy increases. The energy dependence of the cooling time difference perseveres in the presence of energy independent timescales such as the light crossing time. The method is applicable for the brightest {\sl Fermi} $\gamma$-ray flares, and can provide upper limits on the flare location even when the decay times in different energies are statistically indistinguishable.

\acknowledgements
We thank the referee Luigi Foschini for his thoughtful comments and suggestions.

We acknowledge  support from NASA ATFP grant NNX08AG77G, Fermi grant
NNX12AF01G. E.P. and M. G. acknowledge support from LTSA grant NNX07AM17G.

\end{document}